\documentclass[prl,superscriptaddress,twocolumn]{revtex4}
\usepackage[latin1]{inputenc}  
\usepackage[T1]{fontenc}       
\usepackage[austrian,english,french]{babel}
\usepackage{graphicx}
\usepackage{amsmath}
\usepackage{amssymb}
\usepackage{amscd}
\usepackage[sort&compress]{natbib}

\begin{document}
\selectlanguage{english}

\title{Exploiting the randomness of the measurement basis in quantum
cryptography: Secure Quantum Key Growing without Privacy
Amplification}
\author{Hannes R. Böhm}
\email{hannes.boehm@exp.univie.ac.at}
\author{Paul S. Böhm}
\author{Markus Aspelmeyer}
\author{\v{C}aslav Brukner}
\affiliation{Institut für Experimentalphysik, Universität Wien,
Boltzmanngasse 5, 1090 Wien, Austria}
\author{Anton Zeilinger}
\affiliation{Institut für Experimentalphysik, Universität Wien,
Boltzmanngasse 5, 1090 Wien, Austria} \affiliation{Institute for
Quantum Optics and Quantum Information, Austrian Academy of
Sciences, Boltzmanngasse 3, 1090 Wien, Austria}
\date{\today}

\begin{abstract}
We suggest that the randomness of the choices of measurement basis
by Alice and Bob provides an additional important resource for
quantum cryptography. As a specific application, we present a
novel protocol for quantum key distribution (QKD) which enhances
the BB84 scheme by encrypting the information sent over the
classical channel during key sifting. We show that, in the limit
of long keys, this process prevents an eavesdropper from
reproducing the sifting process carried out by the legitimate
users. The inability of the eavesdropper to sift the information
gathered by tapping the quantum channel reduces the amount of
information that an eavesdropper can gain on the sifted key. We
further show that the protocol proposed is self sustaining, and
thus allows the growing of a secret key.
\end{abstract}
\maketitle

\section{Introduction}
Quantum cryptography \cite{gisin02} was first described by Bennett
and Brassard \cite{bennett84} in 1984. Their protocol, commonly
called BB84, is still the most widely used protocol for quantum
cryptography today. Its simplicity, its proven
security\cite{fuchs97} and its possibility to be extended to
entangled photons
\cite{ekert91a,bennett92b,jennewein00,naik00,tittel00} has
contributed to its widespread use.

BB84 describes a protocol for growing a large secret key between
two communicating parties starting from a smaller shared secret.
The processing of the raw data found in BB84 requires several
steps including key sifting, error correction and privacy
amplification. In BB84 privacy amplification \cite{bennett88}
allows to generate a secure key starting from a key that might be
partially known by a possible eavesdropper. This increase in
security comes at the expense of the final key length.

The basic idea of the present paper is to exploit a resource
which, though present in all existing protocols, has sofar not
been utilized to the full extent. The randomness of the basis
choices of both legitimate parties is an important resource as it
is a sequence of perfect random numbers. In current quantum
cryptography protocols it is simply used to choose the basis for
both, preparation and measurement of quantum states. We suggest
that this sequence can be further exploited. For example it might
be used for the encryption of data transmitted between the
legitimate parties.

As an explicit example of the idea, we present here a modification
to the BB84 protocol that reduces the information the eavesdropper
can obtain on the sifted key. It seems that this reduction scales
with the length of the raw key, which would imply that the
information of an eavesdropper on the sifted key can be made
arbitrarily small. Even though a complete security proof is not
given, we suspect that our protocol can work at higher quantum bit
error rate (QBER) compared to privacy amplification at the same
security level. This advantage is gained by further exploiting the
randomness of the measurement basis choices, more than has been
done in BB84.

\section{The Protocol}
Our protocol can be seen as an extension of the BB84 protocol in
the sense that the production of the raw key is identical to the
production in the original BB84. This makes it applicable to both,
QKD based on single photons, as well as entangled state QKD
\cite{ekert91a}. In the analysis of our protocol, we start with
from the original BB84 scheme. However, our protocol can easily be
extended to the case of entangled qubit quantum cryptography, and
probably many other quantum cryptography systems.

The two legitimate communicating parties, called Alice and Bob,
establish a common secret key in the following way. Alice prepares
a state in a two dimensional Hilbert space using one of two
mutually conjugate basis sets and sends it to Bob. In each basis,
one basis vector is attributed to the classical bit value ``0'',
the other to the bit value ``1''. The choice of the basis used,
and the bit value sent, are both assumed to be completely random.

Upon reception of the state, Bob randomly measures the state in
one of the two bases and stores the result together with his
choice of basis used. Now both parties possess a table consisting
of entries for each state transmitted. This table is called the
raw key. Up to this point, our protocol is identical to the BB84
protocol described in \cite{bennett84}.

\begin{figure*}[ht]\vspace{0.1cm}
\center
\includegraphics[width=16 cm, keepaspectratio, clip=true, draft=false]{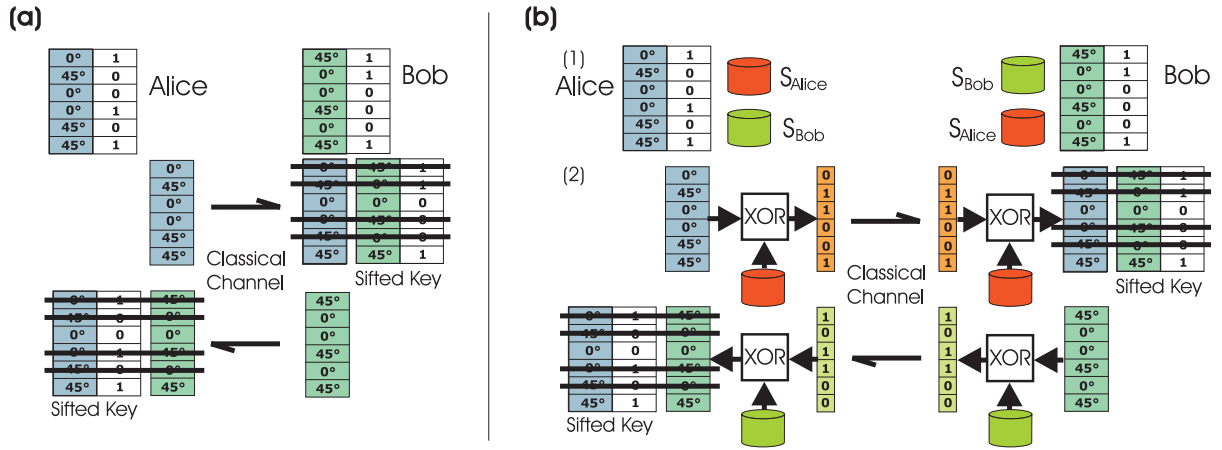}
\caption[The Protocol]{\textbf{(a)} Sketch of the original BB84
sifting method. The bases used to encode and measure the qubits
are transmitted unencrypted over the classical channel. Using the
list of bases received from their respective communication
partners, they can decide which qubits were encoded and
measured in compatible bases and therefore contribute to the
sifted key. \textbf{(b)} Sketch of the protocol proposed in this
paper. (1) Additionally to the lists of BB84, Alice and Bob both
possess a preshared secret that is split into two parts,
$S_{\mathrm{Alice}}$, and $S_{\mathrm{Bob}}$. (2) The information
which basis was used during each individual measurement is
encrypted before it is sent over the classical channel using the
shared secret. This is done by applying a logic XOR between the
list of bases and a part of the shared secret. This encryption of
the encoding and measurement bases renders it impossible for a
third party to correctly sift measurement results obtained from
eavesdropping on the quantum channel. For the protocol to be
secure it is mandatory that Alice and Bob use different parts of
the shared secret and that for successive runs of the protocol,
the secrecy of the shared secret has to be continuously
refreshed.} \label{fig:protocol}
\end{figure*}

Once the raw key is produced it is sifted, which was done in BB84
by publicly announcing the measurement basis on a classical
channel and keeping only the measurement results where Alice and
Bob happened to have chosen the same basis (see Figure
\ref{fig:protocol}a). We have strong indications, that this public
announcement of the measurement bases reveals more information
about the sifted key to an eavesdropper than is necessary for
establishing a secure key between the two legitimate parties. To
overcome this potential weakness of the existing protocols, a
modification of the basis reconciliation process, which does not
publicly announce the measurement basis, is necessary. Note that unlike in
other protocols that omit a public basis announcement
\cite{hwang98}, here the encoding and receiving bases have been chosen
randomly for every transmitted qubit.

Now consider the following situation: The two legitimate communicating
parties, Alice and Bob, have just produced a raw key of length
$n$. Thus, Alice possesses a list containing her random
preparation basis and the random value of the qubit transmitted at
each particular basis choice. Likewise Bob possesses a list
containing his random measurement basis and the corresponding
random measurement result for each qubit received. Every entry in
these lists represent a single transmitted qubit and can be
expressed in two bits of classical information, one for the basis
that has been used ($B_i$), the other ($K_i$) for the prepared bit
value or the outcome of the measurement. Every entry of the list
can thus be written as
\begin{equation}
\left( B_{p,i},K_{p,i} \right)  \,\,\,\,\,
p=\mathrm{Alice},\mathrm{Bob} \,\,\,\,\,i=0,1,\ldots,n
\end{equation}
with $B_{p,i}$ and $K_{p,i}$ being single bit values. Additionally
Alice and Bob share a classical secret $S_{1\dotsc 2n}$. This $2n$
bit secret string has to be available to Alice and Bob before the
protocol starts.  For the further usage it is split into two parts
of equal length $S_{\mathrm{Alice}, {1 \dotsc n}}$ and
$S_{\mathrm{Bob},{1 \dotsc n}}$.

The sifting process now works as follows (see Figure
\ref{fig:protocol}b). Alice and Bob each apply an XOR operation
between their local list $B_{\mathrm{Alice},i}$
($B_{\mathrm{Bob},i}$) and $S_{\mathrm{Alice},i}$
($S_{\mathrm{Bob},i}$) to produce a message $M_{\mathrm{Alice},i}$
($M_{\mathrm{Bob},i}$), in other words
\begin{equation}
M_{p,i} = B_{p,i} \mathrm{\,\,XOR\,\,} S_{p,i}
\end{equation}
The two computed messages ($M_{p,i}$) are then exchanged over a
classical channel. Upon reception, Alice and Bob can decode the
message of their communication partner and regain the original
list of bases by applying the inverse operation
\begin{equation}
B_{p,i} = M_{p,i} \mathrm{\,\,XOR\,\,} S_{p,i}
\end{equation}
After this decoding step, Alice and Bob both have information on
both lists of bases $B_{\mathrm{Alice},i}$ and
$B_{\mathrm{Bob},i}$. Thus, they can now remove all entries of
their record where
\begin{equation}
B_{\mathrm{Alice},i} \ne
B_{\mathrm{Bob},i}\hspace{10pt}\hspace{.1mm}.\label{eqn:sifting-unequal-bases}
\end{equation}
The eavesdropper, called Eve from now on, can not reproduce this
step, because she does not have the shared secret $S_{p,i}$. This
means that even if she has done some sort of eavesdropping on the
quantum channel, her information on the sifted key is less than in
the case of the original BB84 protocol, as she can not correctly
sift the key with certainty.

Once the sifting process is completed, Alice and Bob share a
sifted key, which usually contains errors. In order to generate a
secure key, which can in turn be used for secure transmission of
data, this error has to be estimated and
corrected\cite{brassard94}. After error estimation and error
correction, a secure key is generated from the error-free sifted
key. In BB84 this is done by using a classical privacy
amplification protocol, which washes out the information a
possible eavesdropper could have obtained on the key by
measurements on the quantum channel. In this step the final key is
reduced in length depending on the amount of information an
eavesdropper could possess of the error corrected sifted key.

We will show that in our case the encryption of the classical
channel during basis reconciliation reduces the amount of
information an eavesdropper can get on the sifted key. Even though
we do not have a complete quantitative description of this
reduction of information accessible to Eve, we suspect that the
additional privacy amplification step might not be required under
certain conditions. One has to keep in mind, that the presharing
of a secret string does not represent a disadvantage compared to
BB84, where a shared key is required for authentication of the
classical channel \cite{luetkenhaus99}.

After the secret key has been established between Alice and Bob,
the protocol starts anew with the transmission and measurement of
qubits over the quantum channel. During the new run of the
protocol, the shared secret ($S_{p,i}$) used to encrypt the basis
exchange has to be reused. It is therefore necessary to quantify
the amount of information an eavesdropper can gain about the
shared secret during a single run of the protocol. In general the
upper bound for this information gain depends on the quantum bit
error rate (QBER) as we will discuss in the next section. To
sustain the secrecy of the shared secret it is therefore necessary
to subsequently refresh the secrecy of the initial shared secret
after every run of the protocol.

\section{Security Considerations}
The protocol presented in the last section reduces the possible
knowledge an eavesdropper can obtain on the sifted key that is
established between Alice and Eve. In this section we try to
quantify this reduction of information accessible to the
eavesdropper. In the limiting case of long key length we find a
strong indication that our protocol has an advantage over the
existing combination of BB84 and privacy amplification.
Considering that we use a resource that has not been used to the
full extent in existing protocols, namely the randomness of the
basis choices, it is reasonable that such an advantage exists.

The security analysis is split in two parts. First we analyze the
amount of information about the shared secret that can be
extracted from a single cycle of the protocol. If this amount of
information is smaller than the final sifted key, then it is
possible to grow a longer shared key with our protocol. In the
second part, we analyze the case where Eve has no information on
the shared secret and thus has to sift her measurement results
without any knowledge on the basis used by Alice and Bob.

\subsection{Plaintext Attack}
For the security of the proposed protocol, it is important that
the shared secret can not be determined by analysis of the
messages $M_{\mathrm{Alice},i}$ and $M_{\mathrm{Bob},i}$ and any
resources that are accessible to an eavesdropper.
This includes the ciphertext ($C_i$) that is finally sent by Alice to Bob after
a secret key has been established and full knowledge of the
plaintext ($P_i$) that has been transmitted with this key. These
two resources enable Eve to gain full knowledge of the key that
has been used to send the message. This can be seen by the fact
that, using the Vernam cipher \cite{vernam26}, the ciphertext is
usually created from the plaintext by
\begin{equation}
C_i = P_i\hspace{2mm} \mathrm{XOR}\hspace{2mm}
K_i^{\mathrm{sifted}}
\end{equation}
and thus,
\begin{equation}
P_i = C_i\hspace{2mm} \mathrm{XOR}\hspace{2mm}
K_i^{\mathrm{sifted}}\hspace{2mm}.
\end{equation}

To simplify the further treatment, we assume that Eve has full
information on the raw key. This can be written as
\begin{equation}
B_{\mathrm{Eve},i}=B_{\mathrm{Alice},i} \hspace{.5cm} \mathrm{or}
\hspace{.5cm} B_{\mathrm{Eve},i}=B_{\mathrm{Bob},i} \hspace{.2cm}
\label{eqn:eve-basis-assumption}
\end{equation}
and
\begin{equation}
B_{\mathrm{Bob},i} = B_{\mathrm{Alice},i} \Longrightarrow \\
K_{\mathrm{Eve},i} = K_{\mathrm{Alice},i} =
K_{\mathrm{Bob},i}\hspace{1mm}. \label{eqn:eve-raw-key}
\end{equation}

Note that this assumption provides Eve with more information than
she could obtain with any eavesdropping scheme. For a detailed
security analysis one would have to drop this assumption and
introduce a quantum bit error rate dependent probability for Eve
to have correct bit value for each position in the raw key.
However in our proof of principle analysis is suffices to assume
that Eve has complete knowledge of the raw key.

\begin{figure}[ht]\vspace{0.1cm}
\center
\includegraphics[width=7 cm, keepaspectratio, clip=true, draft=false]{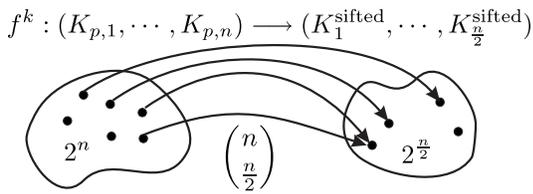}
\caption[Sifting without basis information]{There exist
$\binom{n}{\frac{n}{2}}$ sifting functions that map an $n$-bit raw
key to a $\frac{n}{2}$-bit sifted key. } \label{fig:sifting}
\end{figure}

We now assume that the sifted key consists of exactly half the
number of bits of the raw key, as this is the case with the
highest probability. In this case there exist
$\binom{n}{\frac{n}{2}}$ functions\,\footnote{If the length of the
sifted key is kept secret by Alice and Bob, the assumption that
the sifted key consists of exactly $\frac{n}{2}$ bits gives a
lower estimate for the number of sifting functions.} that
represent a possible sifting method (see Figure
\ref{fig:sifting}):
\begin{multline}
f^{k}: (K_{p,1},\ldots,K_{p,n}) \longrightarrow \
(K_1^{\mathrm{sifted}},\ldots,K_{\frac{n}{2}}^{\mathrm{sifted}} ) \\
k=1,\ldots,\binom{n}{\frac{n}{2}}
\end{multline}
It is easy to see that knowledge on the sifting function $f$ that
was used to create the sifted key, is equivalent to knowledge of
shared secret $S_j$ that has been used during basis
reconciliation.

Without any knowledge of the raw key, thus in the case where Eve
does not extract any information form the quantum channel, Eve can
gain no information about the used sifting function and therefore
about the shared secret.

However, if Eve has maximal information on the raw key as assumed
in (\ref{eqn:eve-basis-assumption}) and (\ref{eqn:eve-raw-key}),
she can try out all possible sifting functions with her own raw
key, and exclude all functions $f^k$ that do not reproduce the
sifted key, she knows from her plaintext analysis. This reduces
the number of possible sifting functions to
\begin{equation}
k_{\mathrm{max}}=\frac{\binom{n}{\frac{n}{2}}}{2^{\frac{n}{2}}}
\sim e^{\alpha n} \hspace{.5cm} 0 < \alpha < 1
\end{equation}
which is still exponentially growing with the key length $n$.

This reduction in the number of sifting functions can be written
as a gain of information $I$ on the shared secret, by calculating
the difference in the Shannon entropy with and without ruling out
the sifting functions that do not reproduce the final sifted key:
\begin{equation}
I = H^{\mathrm{apriori}}-H^{\mathrm{aposteriori}}
\end{equation}
with
\begin{equation}
H = - \sum_i p_i \log_2 p_i \hspace{0.5cm}.
\end{equation}
Assuming that all sifting functions are equally likely,
\begin{equation}
p_i = p \hspace{1cm} \forall p_i
\end{equation}
this reduces to
\begin{equation}
H = - \log_2 p
\end{equation}
and we get an information gain of
\begin{equation}
H = \log_2\binom{n}{\frac{n}{2}} - \log_2
\frac{\binom{n}{\frac{n}{2}}}{2^{\frac{n}{2}}} = \frac{n}{2}
\end{equation}

Because this information gain has been derived for the case where
the eavesdropper has full information on the raw key, this
represents the upper bound on the information an eavesdropper can
gain on the shared secret. To sustain the secrecy of the shared
secret, the legitimate parties have to use this amount of bits
from the generated sifted key to refresh the shared secret. In our
case this would leave the legitimate parties not a single bit for
secret communication. However, this derivation is based on the
unrealistic assumption that Eve possesses full information on the
raw key. One can therefore conclude that the maximal amount of
information on the shared secret in a realistic eavesdropping
scheme is less than the value obtained here. This strongly
suggests that it is possible to use the protocol proposed in this
paper for secure quantum key growing.

Again, we would like to stress that the security analysis
presented is based on
assumption\,(\ref{eqn:eve-basis-assumption}), which gives the
eavesdropper much more information on the raw key than is possible
for any eavesdropping strategy.

\subsection{Sifting without Basis Information}
In the last section we showed that even under the assumption that
Eve has maximal knowledge on the raw key and the transmitted
plaintext, the produced sifted key is sufficiently large to
sustain the secrecy of the shared secret. We now want to show that
the unavailability of the basis information can drastically reduce
the probability to obtain the correct sifted key.

\begin{figure}[ht]\vspace{0.1cm}
\center
\includegraphics[width=7 cm, keepaspectratio, clip=true, draft=false]{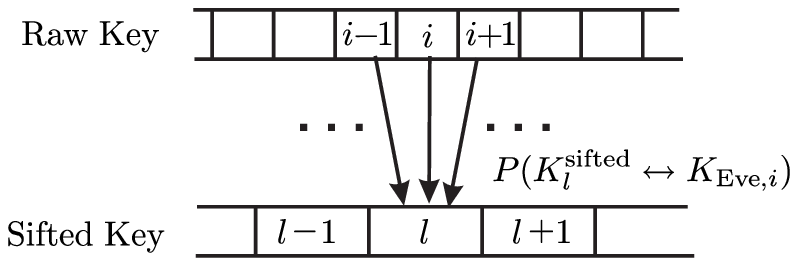}
\caption[]{Without the complete basis information, the
eavesdropper is presented with the following situation: For every
bit of the sifted key, there is a certain probability that it was
derived from a specific bit of the raw key. With increasing key
length $n$, more and more raw key bits contribute with
non-vanishing probability to the specific sifted key bit. This
reduces the probability to conduct a valid sifting process and
therefore reduces the information on the sifted key accessible to
the eavesdropper.}\label{fig:siftcontrib}
\end{figure}

Let us now consider the case where Eve does a simple
\emph{intercept and resend} eavesdropping strategy
\cite{huttner94} on all qubits transmitted from Alice to Bob. If
she uses the same two basis sets as Alice and Bob, the probability
for having a correct final key bit is 75\%, given that she has
full basis information which is needed to sift her measurement
results key. In our protocol, this information is not available to
Eve, and thus a qubit intercepted in a compatible basis does not
necessarily lead to a correct bit in the sifted key. This
reduction in the probability to obtain a correct final key bit
reduces the information accessible to Eve (see Figure
\ref{fig:siftcontrib}).

The probability that a single bit $K_l^{\mathrm{sifted}}$ of the
sifted key is derived from a specific bit $K_i$ in the raw key can
be written by the binomial distribution
\begin{equation}
P(K_l^{\mathrm{sifted}} \leftrightarrow K_{\mathrm{Eve},i}) =
\binom{i-1}{l-1} \left(\frac{1}{2}\right)^i\vspace{.1cm}.
\label{eqn:siftprob}
\end{equation}
For a large number of $l$, this distribution can be approximated
by a Gaussian distribution centered at $i=2l$ and
$\sigma=\frac{\sqrt{i}}{2}$.

The full width of half-maximum of this distribution can be seen as
the number of basis pairs that contribute with a significant
probability to the given sifted key value $K_l^{\mathrm{sifted}}$.
By increasing the length $n$ of the raw key, the information that
Eve can extract from her measurement results can be in principle
reduced arbitrarily.

\subsection{Authentication}
Until now we did not specify the requirements of the classical
channel used in the proposed protocol. One of the important
features of the classical channel in BB84 is message
authentication. There, the authentication of the classical channel
is crucial for the security of the protocol. Without
authentication, a selective modification of the basis
reconciliation process would allow an eavesdropper to decrease the
detectable QBER and thus to hide the quantum error he introduced
during the measurements on the quantum channel.

In our protocol, this selective modification of the basis
reconciliation process is not possible as the bases are encrypted
with the shared secret and therefore completely random. The
plaintext attack does not work to gain information on the shared
secret, because the basis exchange takes place before the transmission of a
ciphertext. However, any modification to a randomly encrypted
message $M_{p,i}$ randomly changes the bases information $B_{p,i}$
and can therefore, in average, not lead to a decreased QBER. This
is an indication, that our protocol could also work without
authentication of the basis reconciliation process. However, until a proof
is found for this argument we have to assume an authenticated classical channel.

\section{Conclusion}
We shown that the random basis choice in quantum cryptography is an important
resource and can exploited more than has been done in existing protocols. Furthermore
we have presented a novel protocol for quantum key growing that
makes use more extensively of the inherent randomness of basis
choices already present in the case of the classical BB84
protocol. By encrypting the information on the classical channel
during the sifting process, it is possible to arbitrarily reduce
the mutual information between a possible eavesdropper and the
legitimate parties. This is due to the fact, that the eavesdropper
can not reproduce the sifting process even in the case where he
has maximal information on the raw key, and partial knowledge of
the final secret key. A complete security proof and comparison
with the full BB84 protocol including error correction and privacy
amplification has still to be constructed. However we suspect that
our method has significant advantages in cases where the QBER is
high and secure bit rates suffer from a heavy decrease due to
privacy amplification.

\section{acknowledgements}
We would like to thank Norbert Lütkenhaus for helpful comments and
Rupert Ursin and Rainer Kaltenbaek for their input in many discussion.
This work was supported by the Austiran Science Foundation (FWF),
Spezialforschungsbereich (SFB) 015 P20, ARC Seibersdorf Research
GmbH (ARCS), by the European Commission, contract no.
IST-2001-38864 RAMBOQ and the Alexander von Humboldt Foundation.

\bibliographystyle{unsrt}
\bibliography{crypto}

\end{document}